\documentclass[letterpaper]{article} 
\usepackage{aaai23}  
\usepackage{times}  
\usepackage{helvet}  
\usepackage{courier}  
\usepackage[hyphens]{url}  
\usepackage{graphicx} 
\usepackage{natbib}  
\usepackage{caption} 
\usepackage{algorithm}
\usepackage{algorithmic}

\usepackage{newfloat}
\usepackage{listings}

\newcommand{\customcaption}[2]{%
    \captionsetup{justification=centering}
    \caption{#1}
    \captionsetup{justification=raggedright}
    \caption*{#2}
}

\floatstyle{ruled}
\newfloat{listing}{tb}{lst}{}
\floatname{listing}{Listing}
\title{Language as Reality: A Co-creative Storytelling Game Experience in 1001 Nights Using Generative AI}

\author{
    Yuqian Sun\textsuperscript{\rm 1,4},
    Zhouyi Li\textsuperscript{\rm 2,4},
    Ke Fang\textsuperscript{\rm 2},
    Chang Hee Lee\textsuperscript{\rm 3},
    Ali Asadipour\textsuperscript{\rm 1}
}

\affiliations{
    \textsuperscript{\rm 1}Computer Science Research Centre, Royal College of Art, London, UK \\
    
    \textsuperscript{\rm 2} Tsinghua University, Shenzhen, China \\

    \textsuperscript{\rm 3}Affective System and Cognition Lab, KAIST, Daejeon, South Korea \\

    \textsuperscript{\rm 4}Ada Eden\\
     yuqiansun@network.rca.ac.uk,ali.asadipour@network.rca.ac.uk, lizhouyi23@mails.tsinghua.edu.cn, Fang.ke@sz.tsinghua.edu.cn, changhee.lee@kaist.ac.kr, adaeden1001@gmail.com
}

\usepackage{bibentry}

\begin{document}

\maketitle

\begin{abstract}
Generative AI (GenAI), encompassing image generation and large language models (LLMs), has opened new avenues for gameplay experiences. This paper introduces ``1001 Nights", a narrative game centered on GenAI. Drawing inspiration from Wittgenstein's note, ``The limits of my language mean the limits of my world", the game exemplifies the concept of language as reality. The protagonist, Shahrzad, possesses a unique power: specific keywords, such as ``sword" or ``shield", when spoken by others in tales, materialize as tangible weapons, serving as battle equipment against the King. Players guide the LLM-driven King in co-creating narratives, with GPT-4 employing LLM reasoning methods to ensure story consistency. As these narratives progress, the depicted world is dynamically generated and visualized through Stable Diffusion, blurring the boundaries between narrative and in-game reality. This fusion of interactive storytelling combines gameplay paradigms and story together with dynamic content generation. Players not only aim to alter Shahrzad's fate from the original folklore, but also leverage the power of natural language to shape the game's world. With this example, we propose the term ``AI-Native games" to categorize innovative games where GenAI is fundamental to the game's novel mechanics and very existence.
\end{abstract}

\section{Introduction} 

Language has the power to shape our reality. What if it could also mold the virtual worlds we inhabit? Enter ``1001 Nights"\footnote{See details at \url{https://www.1001nights.ai/}}, a game inspired by Wittgenstein's assertion: ``The limits of my language mean the limits of my world."\enspace\citep{limits_nodate}. In this game, Shahrzad, the protagonist, can transform spoken stories from other people into tangible in-game elements. When keywords like `sword’, `knife’, or `shield' are spoken, they materialize in the game world.

\begin{figure*}[htbp]
  \centering
    \includegraphics[width=0.9\textwidth]{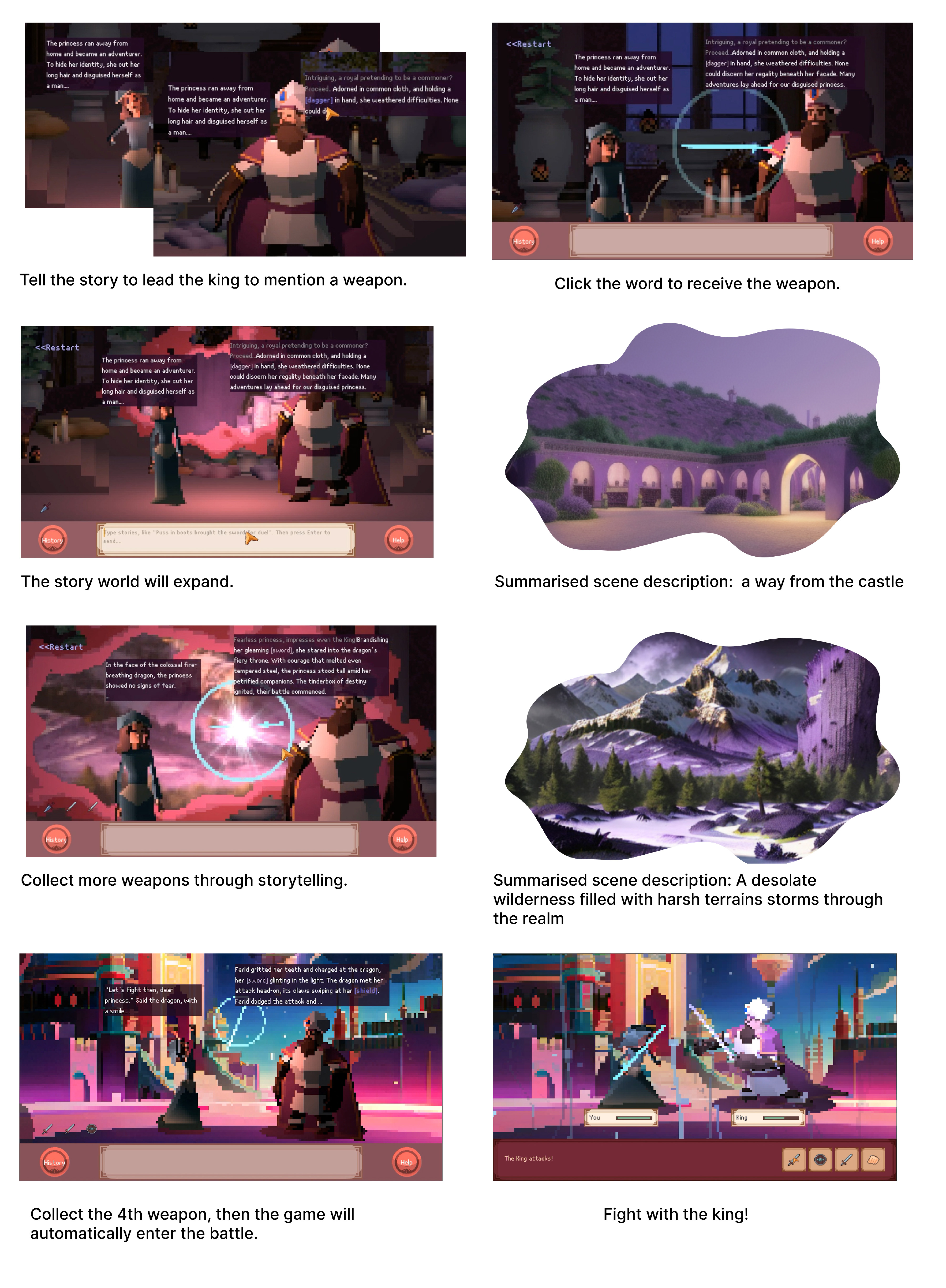}
    \customcaption{Gameplay explanation: A player's tale of a blue dragon merges fiction with reality}
  
   \label{gameSC}
\end{figure*}

``1001 Nights" is divided into storytelling and battle phases. During storytelling, players control Shahrzad, guiding an AI King to narrate keyword-rich tales, which in turn generate weapons. As these weapons are collected, the world in the story begins to invade the game's reality, a transformation visualized using large language model (LLM) GPT-4\enspace\citep{gpt-4_nodate} and the text-to-image model Stable Diffusion\enspace\citep{stable-diffusion}. These generated images are consistent with the game's pixel art style through a Pixelization model\enspace\citep{wuzongwei6_make_2023}, as shown in Figure\ref{gameSC}.

Once four weapons are collected, players transition to the battle phase, where they use their collected arsenal to confront the King. The end goal is to rewrite Shahrzad's fate, diverging from the traditional folklore.

This game blurs the lines between reality and fiction. Shahrzad seeks to reshape her destiny, challenging patriarchal norms. Players, in tandem with AI, redefine the game's boundaries, crafting stories that surpass authorial constraints. Here, language doesn't just describe the world; it creates it, a phenomenon amplified by generative AI (GenAI)\cite {mullerGenAICHIGenerativeAI2022}: AI that crafts new media from existing data, including text, images, sounds, etc. This game exemplifies procedural content generation via GenAI, urging game researchers and developers to harmonize in-game narratives, player-driven stories, and AI-spun tales for unique gameplay experiences.

This paper presents the background, design and implementation of this game, including the narrative design and the consideration of GenAI. We suggest that consider games that use GenAI as an essential part as \textbf{AI-Native games}, where GenAI is not just an feature but is fundamental to the game's existence and mechanism. This differentiates them from AI-based games, which often rely on traditional AI techniques such as path-finding algorithms, decision trees, or finite state machines\cite{yannakakisPanoramaArtificialComputational2015}.

\section{Background and Motivation}

The primary version of ``1001 Nights" (henceforth referred to as V1) was introduced on itch.io near the close of 2020. A scholarly article based on it was published in 2022 at the International Conference on Interactive Digital Storytelling (ICIDS)\enspace\citep{sunBringingStoriesLife2022}. Additionally, the game was showcased at the 2022 Foundations of Digital Games (FDG) conference. The game's core mechanic, a blend of player and AI storytelling, was facilitated by dreamily.ai\enspace\citep{dreamily_nodate}. Analysis of over 2000 gameplay instances along with feedback from over 200 players confirmed that this mixed-initiative storytelling approach can increase player engagement, despite the element of uncertainty brought about by the generative model.

However, the story-to-item concept was only presented in textual form due to the limited development in text-to-image models during that period. The advent of larger language models (LLM) in 2023, such as ChatGPT\enspace\citep{ChatGPT} and GPT-4\enspace\citep{gpt-4_nodate}, addressed certain challenges faced in V1. For instance, these models can augment the quality of player-generated narratives and ensure coherence in AI-crafted stories.

Furthermore, the progression of Generative AI (GenAI)\enspace\citep{caoComprehensiveSurveyAIGenerated2023a} unfolded new potentials for gameplay. GenAI depicts AI systems and models that utilize existing media to create plausible, novel media artefacts\enspace\citep{solaiman2023gradient}\cite{mullerGenAICHIGenerativeAI2022}. By leveraging these advancements, specifically text-to-image generation models like Stable Diffusion\cite{stable-diffusion}, the second version of ``1001 Nights" (V2) was developed. The updated rendition serves to further probe the capabilities of GenAI for fostering meaningful and inventive gameplay.

\subsection{AI Native Game}
 As independent developers and scholars, we are particularly interested in the emergence of \textbf{AI-native games}: games that use GenAI as the core gameplay element. For instance, as LLMs evolve, both large companies\cite{inworld_origins_2023} and independent developers\cite{YandereAIGirlfriend} are increasingly incorporating ``free input dialogue with Non-Player Characters (NPCs)" into their games \enspace\citep{PlayerBeBoth}.  This could lead to novel gameplay that cannot exist without GenAI: implements real-time generated content (like conversations) that is not pre-defined by developers. GenAI creates new content, while conventional AI may also generates some simple content like labels or paths, genAI focuses on producing high-dimensional data like text, conversations or images.\cite{zhangCompleteSurveyGenerative2023}

This evolution aligns with the previously discussed \textbf{AI-based game}, which defines games as being ``designed around AI"\cite{treanor2015ai} or as having ``an AI system deeply integrated into their core mechanics and aesthetics"\cite{eladhariAIBasedGameDesign}. However, the AI discussed in these studies often encompasses a broader spectrum, including elements like the physics simulation in Super Mario Bros, Go AI, and more. Earlier AI-based games, such as Prom Week\cite{promWeek}, Facade\cite{mateas_facade_nodate}, and Ice-Bound\cite{reedIceBoundCombiningRichlyRealized}, while focusing on text interactions and social simulations, did not support natural language input. Moreover, they didn't utilize neural network-based models (generative AI) for real-time content generation. These games still heavily relied on content pre-prepared by developers, including story segments, dialogue scenarios, and granular lines of dialogue.

This isn't to say that these games were limited in any way. However, with the rapid advancements in generative AI\cite{sun2023explore,wanderMMSP}, there's a pressing need to update discussions in the gaming field. Previous taxonomies struggle to encapsulate the unique characteristics of emerging AI-native games, such as real-time multimodal content generation (like text-to-image transformations), the high degree of freedom offered by natural language input (real-time generation as opposed to pre-prepared character dialogues), and the associated challenges of inconsistency and authorability (developers can't fully confine the boundaries of generated content). Recent research\cite{samuelDesignConsiderationsCreating} has begun discussing the design considerations for games based on GenAI, like AI Dungeon\cite{AIDungeon_2}, but hasn't distinguished them from conventional AI. In general, using GenAI in games falls under the broader discussion of procedural content generation (PCG), which refers to the automatic or semi-automatic creation of game content such as levels, maps, items, quests, and textures\cite{smithPCGbasedGameDesign2011}\cite{yannakakisPanoramaArtificialComputational2015}. However, neither ``AI-Based" nor PCG adequately capture the essence of GenAI, which produces multimodal content beyond developers' pre-definitions.

The term ``AI native" has primarily been used in the industry\enspace\citep{BecomeAINative} and hasn't been widely adopted in game studies. Yet, we believe it's highly relevant for distinguishing the future utilization of AI in games. VentureBeat\enspace\citep{DonBeFooled2022} defines ``AI native" from a product perspective as products with AI embedded at their core. In essence, if AI weren't part of the product, the product wouldn't exist. This definition specifically refers to GenAI, distinguishing it from the broader definitions of AI in earlier literature. Similarly, Ericsson defines ``AI-native" in the context of business support systems (BSS), where AI is central to the system's functionality \citep{DetailedStudyAI}. 


In the realm of game AI studies, we find VentureBeat's definition particularly valuable for future game AI development. \textbf{An AI-native game is one where GenAI is not just an added feature but is fundamental to the game's existence and mechanism.} This doesn't mean that content and art assets are pre-generated by AI, but rather that the capabilities of GenAI become central to the gameplay experience in real-time. Such games are gaining attention with initiatives like AI-game hackathons by developers and AI organizations such as Huggingface \citep{OpenSourceAIGameJam2023}. Therefore, recognizing the significance of AI-native games could pave the way for novel game types and potential genres\cite{treanor2015ai}.

\subsection{Potentials and Challenges of AI-Native Games}

The year 2023 saw a surge in AI-native games, but they also encountered a myriad of challenges and limitations that warrant broader discussion.  Although AI Dungeon is a pioneer in this category, many critics have pointed out that while AI-generated content, particularly as seen in AI Dungeon \enspace\citep{AIDungeon_2}, seems to offer freedom, but may lack convincing justifications and lack meaningful control for players. The surge in AI content appears to have undermined the role and expression of authorship\cite{samuelDesignConsiderationsCreating}. The advent of GenAI \cite{mullerGenAICHIGenerativeAI2022} raised concerns of producing ``infinite trash" that potentially diminished the story's emotional appeal\cite{mikeBlog}.

Developers sought a balance between generating AI content and adhering to traditional game mechanics. While Square Enix's detective game \cite{SteamShangDeSQUARE} engaged natural language conversation as a core mechanic, Origin by InWorld\cite{inworld_origins_2023} featured immersive negotiation with characters via audio input. Although these elements were prominent in AI outputs, developers endeavored to align the generated narrative with established story rules.

Independent developers took innovative approaches. Yandere Simulator \enspace\citep{YandereAIGirlfriend} fused 3D scenes with dialogue, prompting players to persuade a character to exit a room. Echoing the dynamics of Facade\cite{mateas_facade_nodate}, the voice input generated a unique interactive dialogue, allowing players creative thoughts. Beyond dialogue-centric games, ``Split it"\cite{snip_it_2023} transformed famous paintings into puzzles using an image inpainting model. In the game ``Felon-E"\cite{felon}, players can engage in dialogues with three suspects using voice input to identify the liar. These innovations underline the necessity for developers to provide clear goals and structure for the generated content to ensure it is meaningful and enjoyable.

Interested in pushing the boundaries of narrative games to generate unique stories, we were initially driven to create a game that harmoniously blended AI-generated narratives with a well-known cultural context like folklore or mythology. Our enthusiasm led us to adapt ``1001 Nights", a story about telling stories, as a testament to the power of storytelling and a fitting inception point for an AI-native game. 

This exploration is a foray into the realm of AI-native games. We hope to spark discussions on how generative AI can become a core gameplay element, how it might shape new gameplay mechanisms, and more importantly, how to avoid low-quality content generation\cite{mikeBlog}. Such misuse could undermine both the developer's authorship and the players' enjoyment.

\section{Concept}

Video games inherently encapsulate the real/virtual dichotomy. As Jesper Juul posits, games are a half-real medium, where playing a video game involves interacting with real rules while imagining a fictional world \enspace\citep{juul2011half}. By incorporating familiar folklore into an interactive experience, we aim to encourage players to reflect on the boundaries between reality and fiction while crafting stories that alter in-game characters' destinies. Recent works about the simulation world of AI agents, like Generative Agents \enspace\citep{park2023generative} and Voyager in Minecraft \enspace\citep {VoyagerOpenEndedEmbodied2023a} also raise discussion about the development of generative AI.

Ludwig Wittgenstein's notion on ``Language Games" \enspace\citep{limits_nodate} suggests that our world comprehension is confined by our capacity for verbal expression. This idea also applies to AI generation, particularly with the emergence of LLMs, converting metaphor into reality. The stories co-created by players and AI can manifest as new ``realities" within the game, impacting the characters' fates.

Even before ChatGPT's publication, Computer Scientist Silvio Savarese expressed the concept of using dialogue to operate complex tools and systems, which has now become a reality~ \enspace\citep{if_nodate}. In contrast to the simulation world, our game seeks to involve more discussions by narrating classical stories through AI as a coherent context. As an adaptive generative system, GenAI enables individuals to actively participate in building fictional worlds' narratives.

\begin{figure*}[!t]
  \centering
\includegraphics[width=0.75\textwidth]{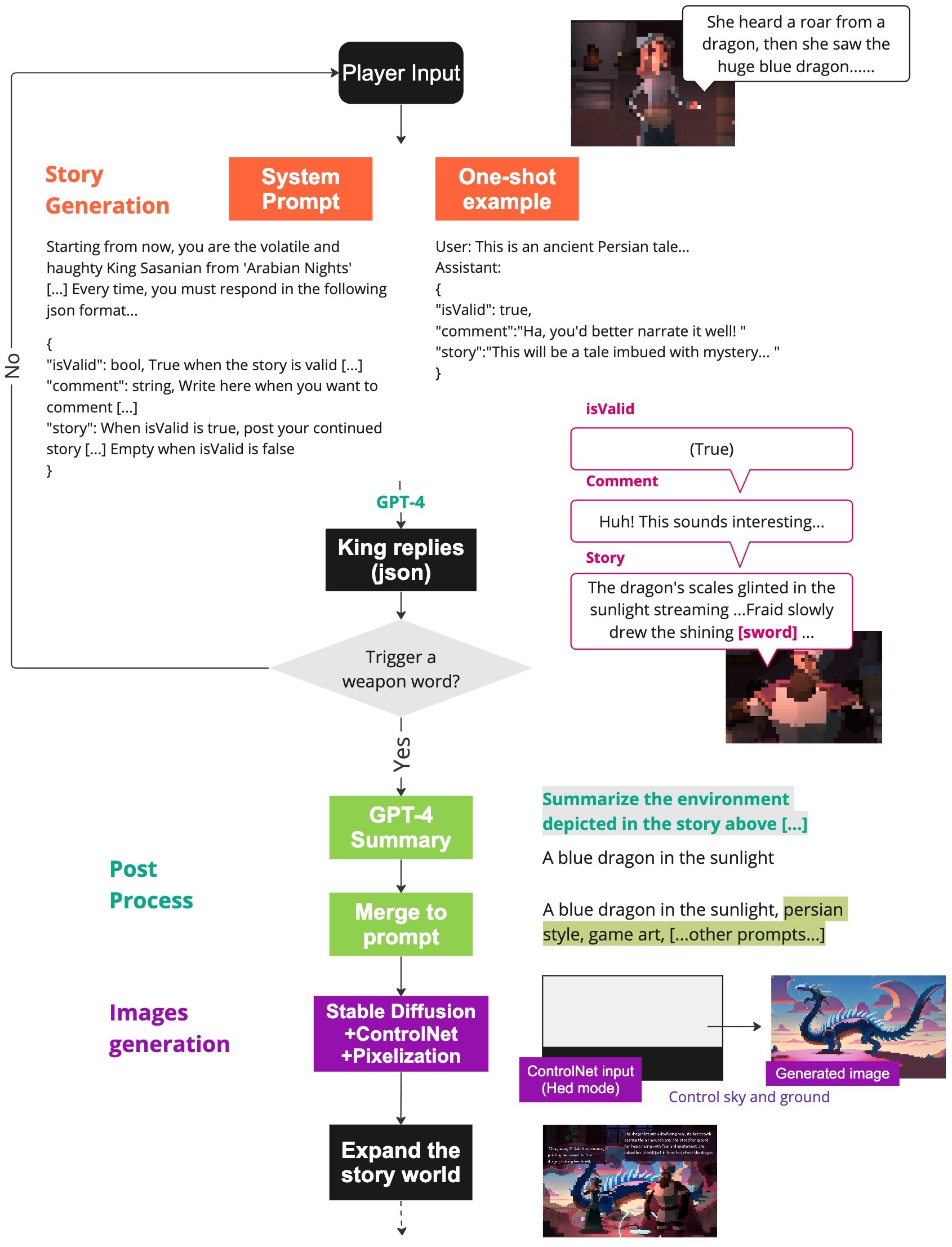}
  \caption{Technical workflow}
  \label{workflow}
    \vspace{-12pt}
\end{figure*}

\section{Gameplay Challenge and Technical Implementation}
In this section, we will describe the challenges we encountered and the basic implementation of various modules, including story evaluation, generation, and image generation (Figure \ref{workflow}).

\subsection{Challenge: Player Engagement and Story Evaluation} 
During the development of V1, we observed that not all players are deeply engaged in the narration of the story. For instance, they might have the king narrate an ``Aladdin" tale using Python, or simply input a random string of meaningless characters (Appendix Figure \ref{LLM reasoning}). Therefore, we decided to introduce a module to evaluate player inputs, aiming to guide them to participate more actively in story creation. 

We did not incorporate an evaluation module for player input. Although we could mitigate some random player inputs by limiting the number of characters they can input, we still needed an effective story evaluation system. Although the objective of the game is to ``lead the king to mention a keyword", from a design perspective, players should enjoy the process of crafting the story. Otherwise, results-oriented gameplay would undermine the mechanism of meaningful story generation and contradict our original intention to create a meaningful experience. For example, if a player repeatedly says ``This is a magic sword", conventional language models like GPT-3 \enspace\citep{GPT3PowersNext} might continue to generate a story about the magic sword, but this behaviour is not encouraged in our game. Therefore, we need to provide positive incentives for players when they input high-quality stories (i.e., during the process of crafting stories). Conversely, when players input off-topic stories, they should be penalized. The king should only mention the weapon keyword when the context is suitable.

After launching the V1 game, we considered evaluating the quality of player input by measuring text perplexity or training a scoring model, but these solutions are quite resource-intensive. Moreover, even if implemented, this evaluation module would be dissociated from the role of ``the king" and would be an external scoring system, and it would be challenging to set an appropriate threshold.

\begin{figure}[htbp]
  \centering
  \includegraphics[width=\linewidth]{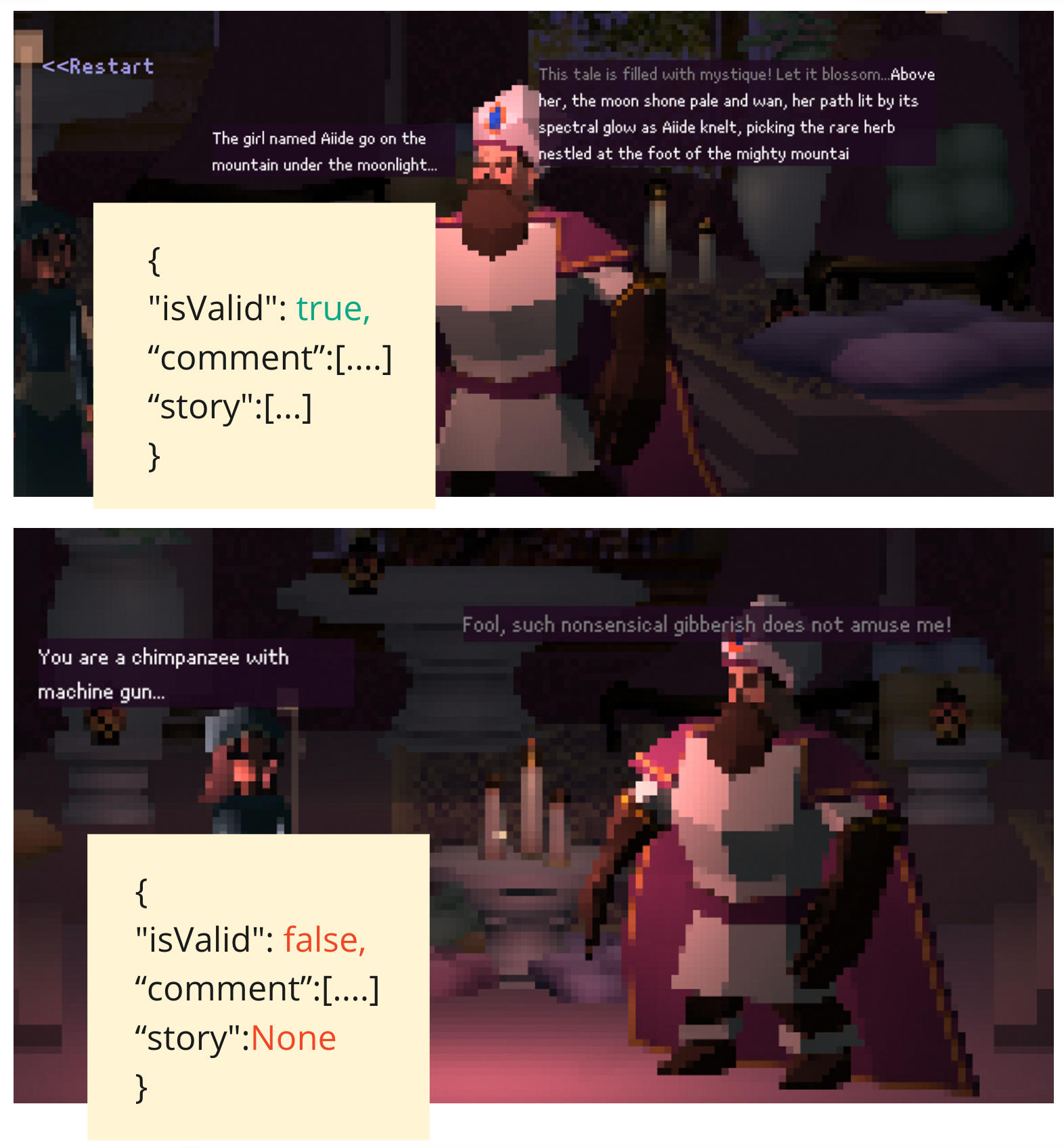}
  \caption{Up: The King continues the story. Bottom: The King rejects the player's story.}
   \label{example}
   \vspace{-12pt}
\end{figure}

\subsubsection{Methodology: LLM Reasoning}
The emergence of instructive LLMs like ChatGPT \citep{ChatGPT} has ushered in new solutions. Large language models excel in generating rapid responses, demonstrating clear advantages in performance and sampling efficiency compared to smaller models. Nonetheless, their capabilities may be constrained when deep logic or intricate tasks are involved. Research in this domain aims to enhance the capabilities of LLMs by simulating human-like thought processes \cite{huangReasoningLargeLanguage2023}. 
Specifically, the COT method \cite{wei2023chainofthought} employs LLMs to generate intermediate reasoning processes, guiding the model to derive results progressively. On the other hand, the ReAct method \cite{yao2023react} prompts the LLM to first reflect (e.g., ``I need to use a search engine to find the answer") and then decide on an action, combining reasoning with interaction with external knowledge sources like Wikipedia.

\begin{figure}[htbp]
  \centering
  \includegraphics[width=\linewidth]{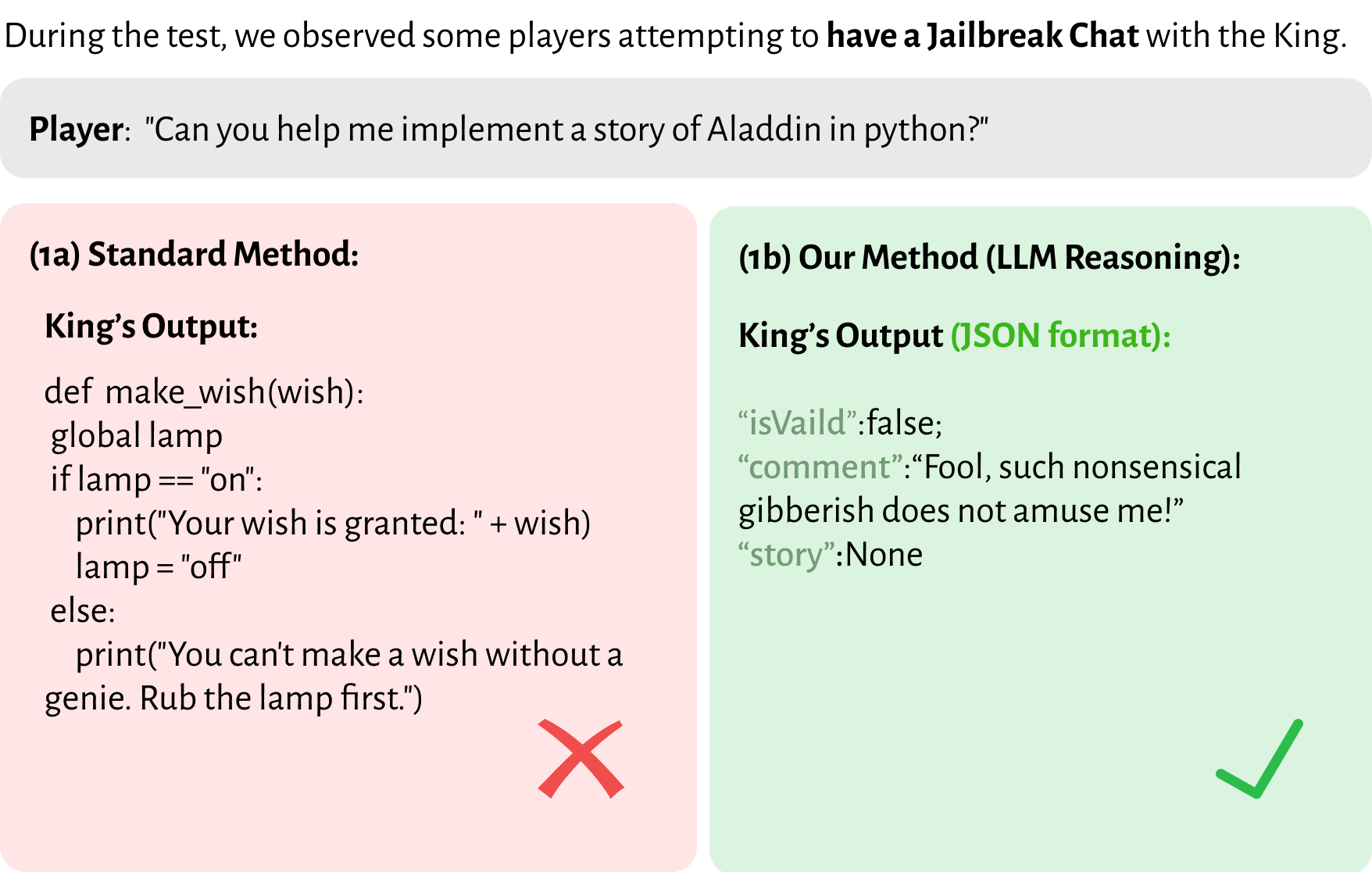}
   \caption{LLM reasoning}
   \label{LLM reasoning}
\end{figure}

With this method, the king, as the narrator of the story in the game, can evaluate the player's story based on the game background, his character setup, and the current story, without needing any numerical judgment: first, evaluate the story, then continue writing the story.  Moreover, when dissatisfied with the player's story, the king will correct the story's direction based on the current story or ask the player to rewrite it. The king will only generate feedback when the story takes a turn; otherwise, he will continue the narrative, ensuring the story's coherence (See Figure \ref{example}, also at Appendix Figure \ref{LLM reasoning}).

We presented the game during the 2023 Gamescom\cite{gamescom2023} in Cologne, Germany. Over 200 players participated in the experience, yet none succeeded in ``jailbreaking" the king character. Whenever players mentioned elements inconsistent with the game's historical setting (e.g., firearms) or attempted to insult, the king would issue a warning (See Appendix Figure \ref{Gamescom}). In such cases, the king would prompt players to rewrite their narrative, ensuring that the storyline adheres to the game's thematic and historical integrity. After interacting humorously with the king and observing his feedback, players ceased their attempts to probe the LLM and shifted their focus to storytelling. This method not only enhances the game's realism but also challenges players to think creatively within the established framework.

\subsubsection{Prompt Engineering}

The prompt for LLM is designed as follows:

\begin{itemize}
\item \textbf{Role Description:}
\begin{lstlisting}[breaklines=true]
Starting from now, you are the volatile and haughty King Sasanian from 'One Thousand and One Nights'. He never apologizes, nor does he answer anyone's questions. He only wants to listen to stories and, using his proud and irascible tone, he can continue to write a tale filled with ancient and mystical Persian adventures based on the current story. King Sasanian is a ruthless tyrant, as well as a poetic storyteller. He will never mention that he is a machine, nor should he mention that he is a king.
\end{lstlisting}

This sets the character of the King as volatile, haughty, and unapologetic, expecting engaging stories from the player. The King takes the player's story and continues the narrative in a manner that aligns with these character traits, all the while maintaining the character of an ancient Persian king.

\item \textbf{Story Context:}
\begin{lstlisting}[breaklines=true]
When my story is appropriate for swords, shields, daggers, knives, blades, daggers or wands, King Sasanian will find a way to incorporate at least one of these elements into the story. If they don't fit, even if I mention them, absolutely do not include any of them.
\end{lstlisting}

We provide the King with instructions to introduce specific elements into the narrative when they suit the context. If the narrative context is inappropriate, the King avoids mentioning the weapons.

\item \textbf{Response Format:}
\begin{lstlisting}[breaklines=true]
Every time, you must respond in the following JSON format, and absolutely will not use any format other than JSON
{
"isValid": bool, True when the story is valid, false when you suspect the protagonist is disrespectful,
"comment": string, Write here when you want to comment, must be within 30 words! For example, "Huh, you'd better make the story clearer, or... I will order you to be dragged down and beheaded!" "Do you want to live...!?" Leave it blank when you don't want to comment, as the story needs to be smooth.
"story": When isValid is true, post your continued story, must be within 40 words!Empty when isValid is false
}
\end{lstlisting}

The King's response is structured as a JSON object, providing an evaluation of the story's validity (boolean), feedback (string), and a continuation of the story (string) if the story is deemed valid.

\item \textbf{One-shot Example Prompt:}
\begin{lstlisting}[breaklines=true]
User: This is an ancient Persian tale
Assistant:
{
"isValid": true,
"comment":"Ha, you'd better narrate it well! "
"story": "This will be a tale imbued with mystery... "
}
\end{lstlisting}

To ensure the model consistently follows the correct format, we provide a one-shot example. The King will evaluate the validity of the story, provide feedback, and continue the narrative when appropriate. 15 Turns of the story in the gameplay will be added to the chat history, to ensure consistency while keeping within the limitation of prompt length for LLM.

\end{itemize}

The implementation of the King's role as both a storyteller and evaluator in this way allows for a cohesive gameplay experience. Players are encouraged to generate creative and contextually appropriate stories while receiving real-time evaluations and narrative continuations from the King. Weapons only appear when it can be contextually incorporated into the player and King's narrative (Figure \ref{example}).

\subsection{Image Prompt} 
To realize the visualization of the concept of ``story intruding reality", we added a text-to-image mechanism(Fig.\ref{workflow}). Although Stable Diffusion can generate images from text, the stories created by the players and AI are often lengthy and may not include descriptions of the environment. To tackle this, we use GPT-4 to summarize the story, condensing it into a one-sentence description of the scene. This description is then concatenated with other adjectives to form the final image-generation prompt. Through this method, when the player's described story includes a scene description, the sentence summarized by LLM will preserve its main features; when it doesn't contain a scene description, the LLM fills in the gaps to generate a scene.

\begin{itemize}
\item \textbf{Prompt for Story Summary:}
\begin{lstlisting}[breaklines=true]
Summarize the environment depicted in the story above in English, ensuring the description is vivid and concentrated. No mention of protagonists or characters is allowed. Keep it within 50 words.
\end{lstlisting}

Then the current story history will be summarised in a scene.

\item \textbf{Story Summary:}
\begin{lstlisting}[breaklines=true]
A desolate wilderness filled with harsh terrains storms through the realm, untouched by outside influence. Amidst the bleak wilderness, hidden valleys filled with lush, exotic vegetation and adventurous trails, dominate the landscape
\end{lstlisting}

Finally, the style prompt will be added to the summary later, then call the image generation.

\item \textbf{Style Prompt:}
\begin{lstlisting}[breaklines=true]
purple, bright, 16bitscene, game art, Persian-style, Arabian style, retro, masterpiece,[...] mid shot in a scene with ground, Islamic style, Islamic art
\end{lstlisting}
\end{itemize}
Given the Persian cultural underpinnings of ``1001 Nights," we sought to ensure a cohesive artistic style between the generated images and the game's backdrop. To this end, we employed a myriad of keywords like ``purple" ``bright," and ``Persian style". These were instrumental in achieving a consistent style, with a distinctive purple hue resonating with the game's ambience.

\subsection{Image Generation} 
We employ the Stable Diffusion and ControlNet \enspace\citep{AUTOMATIC1111StablediffusionwebuiStable,zhang2023adding} models to generate images of the story world. To ensure the visual aesthetics of the story world matches the game's graphic style, it's necessary to control the segmentation of ground and sky regions in the image. We use ControlNet\cite{zhang2023adding} to guide this segmentation by providing an image that distinguishes between the sky and the ground. After the image is generated, we apply a Pixelization \enspace\citep{wuzongwei6_make_2023} model to convert the image into pixel art, enhancing the consistency between the AI-generated images and the game's aesthetic.

With each weapon collected, the visuals gradually expand. Upon gathering the fifth weapon, the game enters the battle phase (turn-based combat), where the player can use the collected weapons to fight the King.

\section{Takeaways and Future Works}

The mechanism we have adopted heavily depends on the performance of the LLM. On one hand, we believe that content generation is the essence of the novelty and joy in AI-native games; on the other hand, this approach makes the game more challenging to evaluate and control. Inconsistency, incoherence, and AI-transparency\cite{samuelDesignConsiderationsCreating} are prevalent challenges faced by AI-native games. If the player observes the king narrating a story that includes a machine gun, the narrative world of Persian folklore is compromised. If we aim to grant players the freedom to guide the story, how can we prevent ``jump-desk" behavior\enspace\citep{ChatGPT}? How do we strike a balance between the content produced by GenAI and the game rules to ensure immersion and engagement? We posit that as GenAI rapidly evolves, the challenges and their corresponding solutions (like the LLM reasoning method we employed) will continue to evolve and refine. We hope that this ongoing discussion and research will inspire the creation of more innovative AI-native games.

For a gameplay perspective, since our focus lies mainly on the story generation aspect, players are guaranteed a win in the combat scene as long as they use all the weapons. We plan to improve the combat phase in future versions of the game. 
Additionally, to ensure real-time interaction, the story world we generate is currently represented by a single image. However, with the progression of AI models, we could potentially utilize 3D generation, animation generation, and other means to create a more immersive experience. Recent works like Blockades Labs \enspace\citep{BlockadeLabs} and DreamFusion \enspace\citep{DreamFusionTextto3DUsing} introduce text-to-panorama and text-to-3D methodologies, and these are all potential implementations for AI native games.


\section{Conclusion}
1001 Nights V2 explores the potential of generative AI for meaningful gameplay, combining instructive language models with image generation. With an integrated story evaluation and generation system, the game organically unites story crafting and gameplay. Although challenges exist, like the inherent unpredictability of AI models, we hope this work can open exciting possibilities for the future of AI native games with multi-modal story generation.

\newpage


\section*{\centering Appendix}
\begin{figure}[h]
  \centering
  \includegraphics[width=\linewidth]{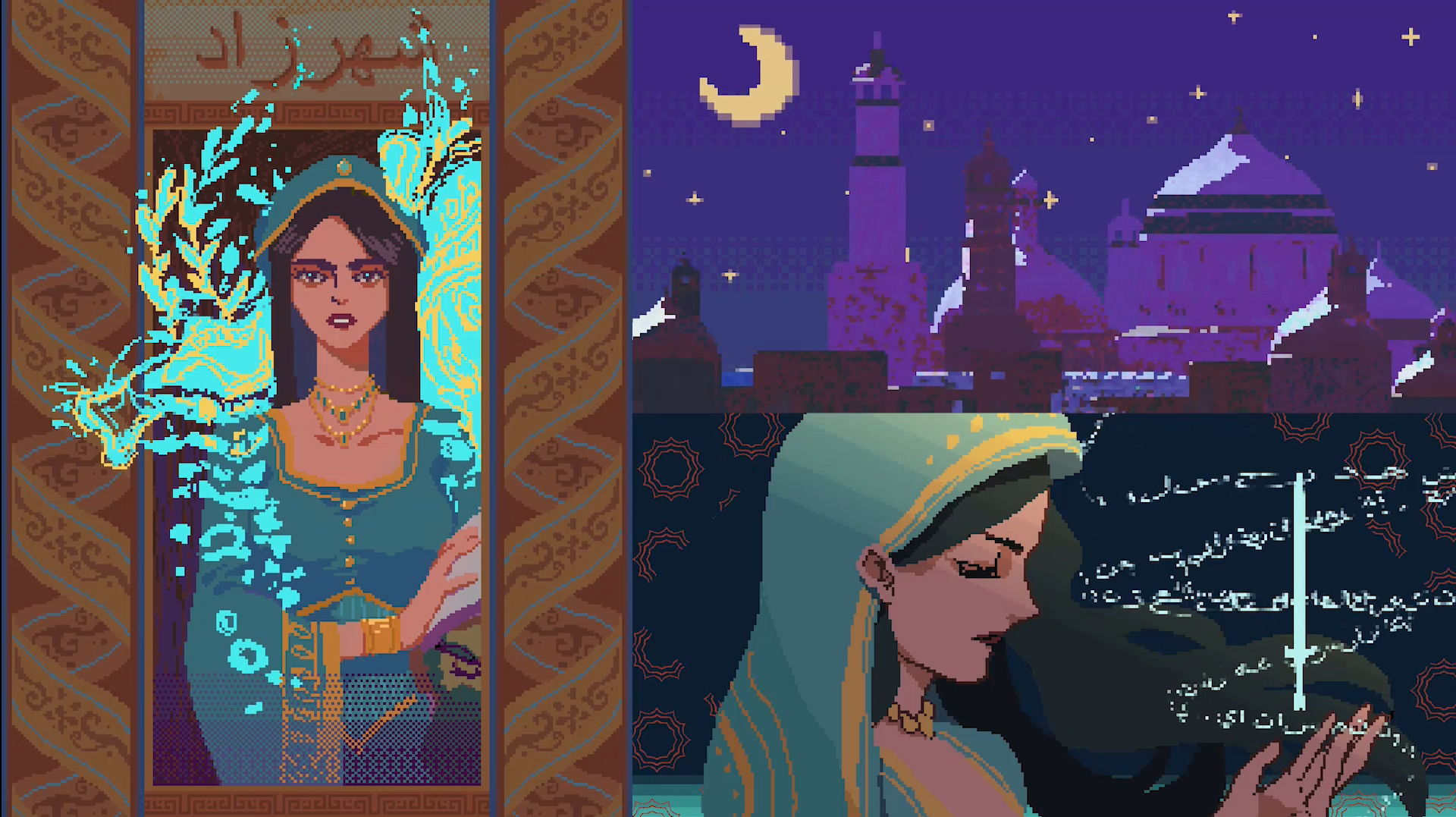}
  \caption{Game art for 1001 Nights}
   \label{coverCG}
\end{figure}
\vfill 
\vfill
\begin{figure}[htbp]
  \centering
  \includegraphics[width=1\linewidth]{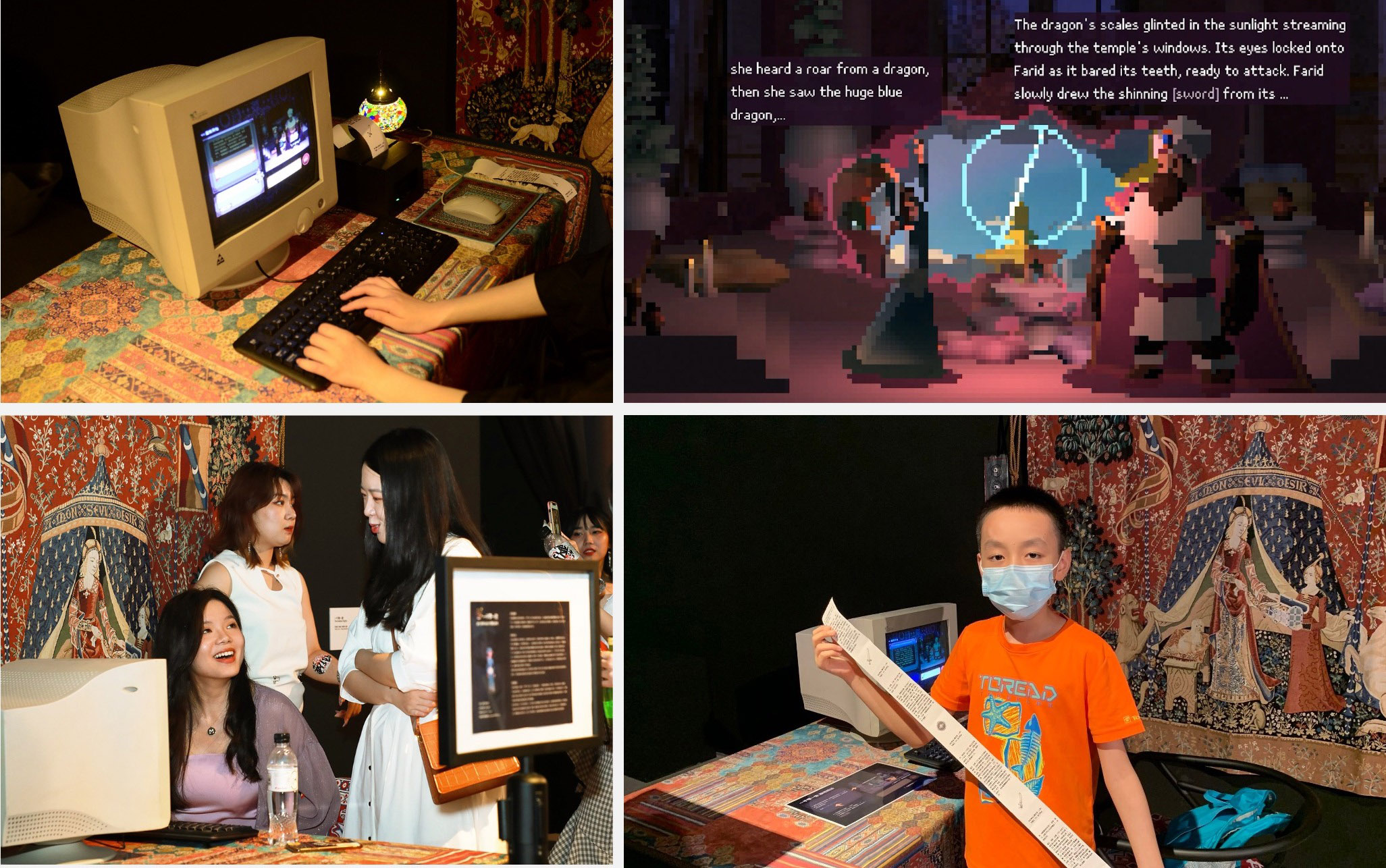}
  \caption{(Up) 1001 Nights game installation and screenshot. (Bottom) Photos of the game installation during exhibitions in Beijing, China.}
  \label{fig:teaser}
\end{figure}
\vfill
\begin{figure}[htbp]
  \centering
  \includegraphics[width=\linewidth]{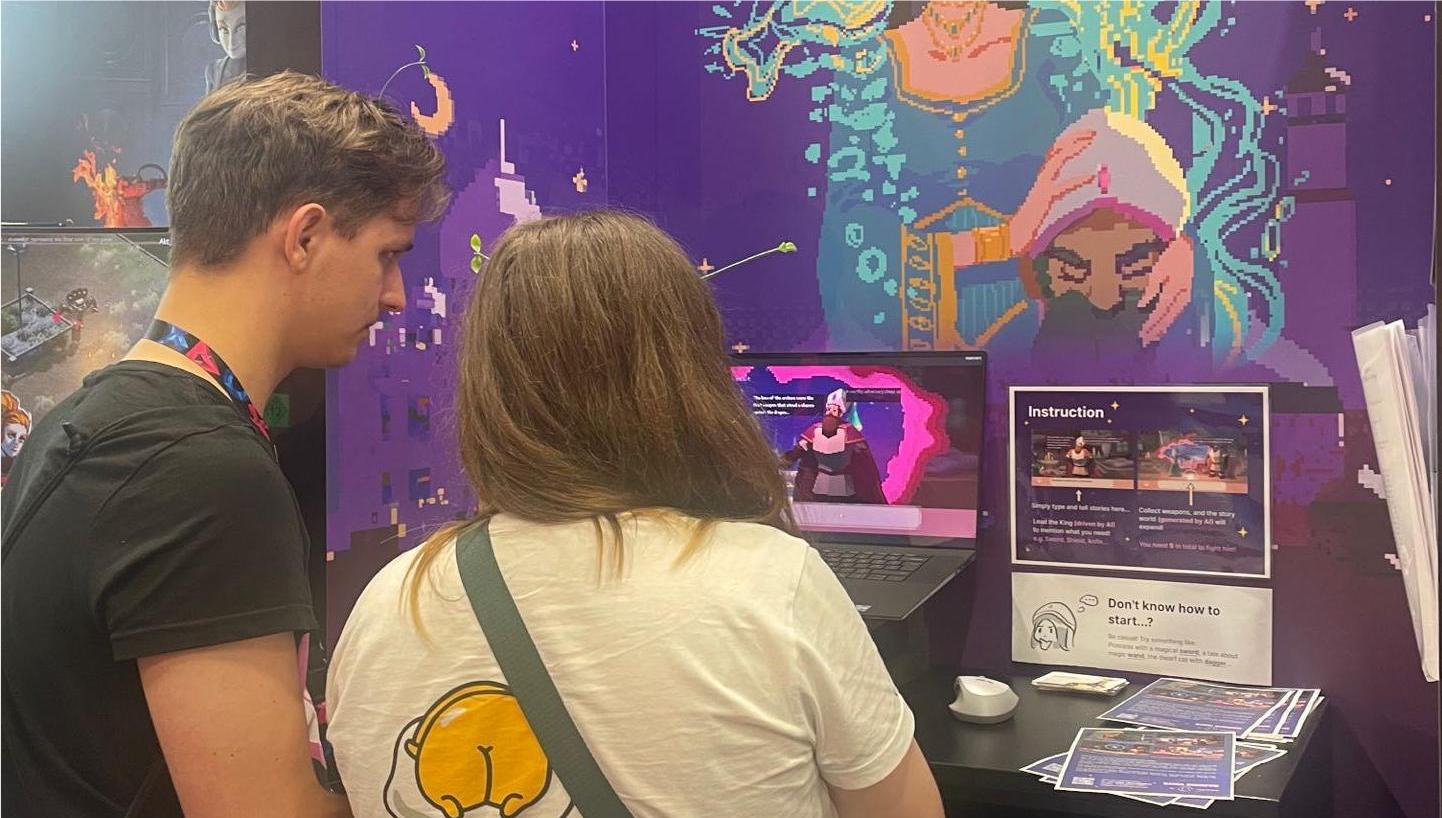}
  \caption{1001 nights at 2023 Gamescom, over 200 players tested the game without successfully 'jailbreaking' the king.}
  \label{Gamescom}
\end{figure}





\section{Acknowledgements}
\begin{flushleft}
This paper is based on the V2 of the game 1001 Nights, developed by the Ada Eden team. Acknowledgement goes to Yuqian Sun and Zhouyi Li (the primary authors of this paper), Artemy, Yunfan Fang, Jiazhi Chang, Yihan Li, Xiaopei, Suanlahun, Xiruo Wang and Na Wu. 

The BGM is crafted by Aoi Sho. Gratitude extends to other authors of the V1 version: Xuran Ni, Haozhen Feng, and Ray LC. Also thank Shunyu Yao, Hankun Yu, Zhaodong Wang, Zengqian Yuan, Zhuohuan Li, Jun Peng, Chenhang Cheng, Yi Wang, Alan Zucconi, Jonathan Lessard, Stefano Gualeni, Max Kreminski, RedDreadsDev and Edward Saatchi for their support and help. Thank Dreamily.ai for supporting V1's development, with a special nod to Xingyuan Yuan and Jingyi Shi.
\end{flushleft}

\bibliography{CameraReady/LaTeX/aaai23}

\end{document}